\documentclass{ws-p8-50x6-00}

\begin{document}

\title{RELATIONSHIP OF THE $^3P_0$ DECAY MODEL \\ TO OTHER STRONG DECAY MODELS}

\author{B. Desplanques, A. Nicolet and L. Theussl}

\address{Institut des Sciences Nucl\'eaires (UMR CNRS/IN2P3-UJF),  \\
 F-38026 Grenoble Cedex, France \\ 
E-mail: desplanq@isn.in2p3.fr}


\maketitle

\abstracts{
The  $^3P_0$ decay model is briefly reviewed. Possible improvements, 
partly motivated by the examination of a microscopic description of a 
quark--anti-quark pair creation, are considered. They can provide support 
for the one-body character of the model which, otherwise, is difficult to 
justify. To some extent, they point to a boost effect that most 
descriptions of processes involving a pair creation cannot account for.}

\section{Introduction}
The $^3P_0$ decay model, first introduced by Micu\cite{Micu:1969mk}, has been 
subsequently applied to the description of many processes by Le Yaouanc 
et al.\cite{LeYaouanc:1988}. Since then, it has been used extensively with a 
reasonable success, especially for the hadronic decays of mesons. Being 
described by a one-body operator, the model can be employed easily, while 
its strength is generally fitted to experiment. In these conditions, the 
agreement is not much better than a factor 2, which is too large to make 
stringent tests of the description of hadrons for instance. Improvements 
should therefore be introduced. This however requires to understand what 
the model accounts for. 
After reviewing the model, we will consider possible improvements, based 
on a microscopic description of a quark--anti-quark pair creation.

\section{The $^3P_0$ decay  model}
\begin{figure}[htb]
\begin{center}
\mbox{\psfig{file=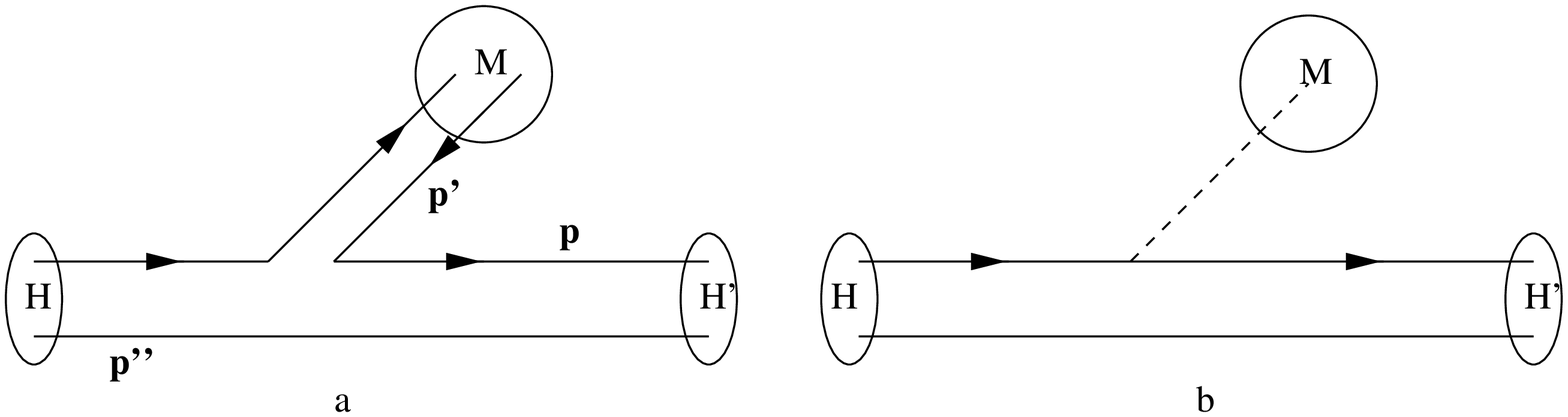,width=\textwidth}}
\caption{Diagrams representing a meson decay with a one-body interaction: 
within the $^3P_0$ decay  model (a) and an elementary emission model (b).  
\label{fig:1body}}
\end{center}
\end{figure}
The $^3P_0$ decay model assumes a creation of a quark--anti-quark pair from the 
vacuum with the corresponding quantum numbers, $J=0$, $L=1$, $S=1$, $T=0$. 
Represented 
in Fig.~\ref{fig:1body}a for a meson decay, it may be described by the following operator:
\begin{eqnarray}
H_{pair}=\gamma \;\sum_{i,j} 
a_{i}^{\dagger}(\vec{p}\,)  \; b_{j}^{\dagger}(\vec{p}\,') \; 
\frac{ \vec{\sigma} \cdot (\vec{p}-\vec{p}\,') }{2\,\sqrt{2\,\pi}} \; 
(2\pi)^3\; \delta(\vec{p}+\vec{p}\,')+h.c. 
\label{3P0}
\end{eqnarray}
In comparison with the elementary emission model shown in Fig.~\ref{fig:1body}b, where 
a meson is emitted from a quark line, it offers the considerable conceptual 
advantage that all hadrons are considered on the same footing. 

Among improvements, it has been proposed\cite{Roberts:1998kq} to introduce 
some momentum dependence in the strength $\gamma$. A relativized 
version of the model has also been considered, involving the replacement 
of $\vec{\sigma} \cdot (\vec{p}-\vec{p}\,') $ in Eq.~(\ref{3P0}) by $ 
2\, m\;\bar{u}(\vec{p}\,) \; v(\vec{p}\,')$\cite{Cano:1997cq}. Some improvement 
is obtained but in absence of insight on the origin of the model, 
one cannot draw firm conclusions. 

\begin{figure}[htb]
\begin{center}
\mbox{\psfig{file=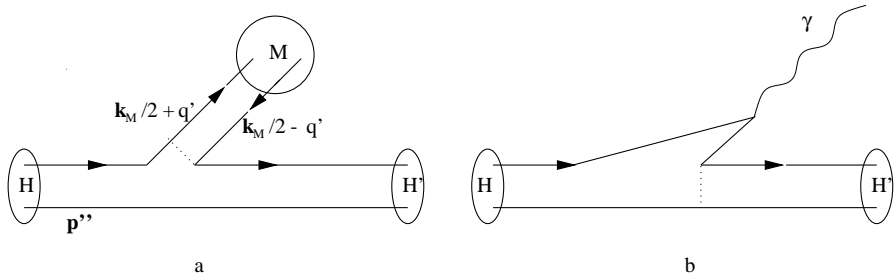,width=\textwidth}}
\caption{Diagrams representing a meson decay with a two-body interaction: 
emission of a meson (a) and a photon (b).  \label{fig:2body}}
\end{center}
\end{figure}

Although it should be used in the c.m. of the decaying system, the creation 
of a pair from the vacuum without relation to the other particles is 
difficult to imagine. Moreover, a term like Eq.~(\ref{3P0}) can be absorbed 
in a redefinition of the quarks and their masses, producing elementary 
quark-meson couplings for instance. Kokosky and Isgur discussed the 
possibility that the $^3P_0$ decay model could account for a flux-tube 
breaking in some limit\cite{Kokoski:1987is}. The strength is not known however. 
Another possibility is that the pair creation is closely related to the 
interaction between quarks\cite{Ackleh:1996yt}. This hypothesis, that we will 
develop in the following, is illustrated in  Fig.~\ref{fig:2body}a for a meson decay. 
%

\section{A key relationship}
The calculation of the decay amplitude of a hadron $H$ into another one $H'$ 
and a meson $M$, as depicted in Fig.~\ref{fig:2body}a for a particular case, implies an 
expression like the following:
\begin{equation}
A_{(H \rightarrow H'+M)} \propto \int \frac{d\vec{p}\,''}{(2\,\pi)^3} \; 
\psi_{H}^{*}(\dots) \;  \psi_{H'}(\dots) \;
\int \frac{d\vec{q}\,'}{(2\,\pi)^3} \;V(\vec{q},\vec{q}\,') 
\;\phi_{M}(\vec{q}\,'),
\label{amp2}
\end{equation}
where $\phi_{M}(\vec{q}\,)$ represents the meson wave function. 
$V(\vec{q},\vec{q}\,')$ represents the interaction responsible for a 
pair creation. Quite generally, its expression is complicated and 
without relation to the interaction between quarks. In 
one case, however, such a relation can be established. For 
spin-less particles, both interactions are the same (non-relativistic 
approximation). One can then use the equation that the meson wave 
function has to fulfill:
\begin{equation}
\int \frac{d\vec{q}\,'}{(2\,\pi)^3} \;V(\vec{q},\vec{q}\,') 
\;\phi_{M}(\vec{q}\,') = (E-2\,e_q) \; \phi_{M}(\vec{q}\,).
\label{key}
\end{equation}
This can be used to replace the last factor in Eq.~(\ref{amp2}) by 
the wave function itself, obtaining:
\begin{equation}
A_{(H \rightarrow H'+A)} \propto \int \frac{d\vec{p}\,''}{(2\,\pi)^3} \; 
\psi_{H}^{*}(\dots) \;  \psi_{H'}(\dots) \;
``(E-2\,e_q)" \; \;\phi_{A}(\vec{q}\,),
\label{amp1}
\end{equation}
where $\vec{q}\,$ has to be replaced appropriately in terms of the external 
momenta and $\vec{p}\,''$.
This last expression looks very much like the one used when employing the 
$^3P_0$ decay model (spin put apart). It does not contain the interaction 
explicitly but involves a one-body operator. This one however appears 
with a well defined factor, $``E-2\,e_q"$, providing a clue for both the 
strength and the momentum dependence, often introduced on a 
phenomenological basis. 

The expression of the above amplitude corresponds to a  
particle--anti-particle pair creation that has the same form as the one 
appearing in a free-particle interaction. It however differs by essential 
features. As already mentioned, such a term can be absorbed into the 
redefinition of the particle fields. The functional dependence of the front 
factor, $``E-2\,e_q"$, rules out this transformation here, since it 
cancels for free particles, consistently with the fact that it originates 
from an interaction term. This can solve one of the problems with the
$^3P_0$ decay model, namely the creation of a pair without relation 
to the environment.

With the above respect, it is worthwhile to mention another example 
where an interaction effect can be turned into a single-particle 
contribution. It concerns the Z-type contribution to the emission of a 
photon shown in Fig.~\ref{fig:2body}b. This one evidently involves the interaction. 
However when the full Feynman diagram corresponding to this figure is 
considered, it is found that the same contribution appears as a 
single-particle one. Consistency is achieved by the fact that 
this contribution involves a factor similar to the above one, $``E-2\,e_q"$. 
The details can be checked by looking at a simple 
model\cite{Desplanques:2001ze}.

When applying the above ideas to spin $1/2$ particles, one has to take 
into account that many terms of the order $\vec{\sigma} \cdot \vec{p}$ are 
produced by the interaction, besides those appearing at the vertex 
where a quark--anti-quark pair is created. While the latter ones can be 
accounted for by the $^3P_0$ decay model, the other ones cannot. For 
that particular contribution, which could represent some average one, 
the structure of the operator is obtained as above for spin-less 
particles and is given by the replacement in Eq.~(\ref{3P0}):
\begin{equation}
\gamma\; \frac{ \vec{\sigma} \cdot (\vec{p}-\vec{p}\,') }{2\,\sqrt{2\,\pi}} 
\rightarrow ``(E-2\,e_q)" \; \bar{u}(\vec{p}\,) \; v(\vec{p}\,').
\label{repl}
\end{equation}
If the meson is strongly bound ($E$ negligible) and $e_q$ approximated 
by the quark mass, the value  $2\sqrt{2\,\pi} \simeq 5$ is obtained for 
$\gamma$. This 
value compares well to the value that can be obtained from meson 
decays\cite{Ackleh:1996yt} but is significantly smaller than the one derived 
from baryon decays\cite{Cano:1997cq,Theussl:2000sj}. 
Within conventions, the sign that is required in some cases is also known. 

A relativistic approach based on the description of mesons by 
Bethe-Salpeter amplitudes suggested some similarity with the quark-pair 
creation model\cite{Kulshreshtha:1983zu}. Comparison with the present work 
indicates a close relation which strongly supports the hypothesis that 
the $^3P_0$ decay model, as completed here, would implement boost 
effects that an approximate (non-relativistic) description of hadrons 
cannot account for. The possibility that it still accounts for other 
effects remains open however.

\end{document}